\newskip\humongous \humongous=0pt plus 1000pt minus 1000pt
\def\caja{\mathsurround=0pt}
\def\eqalign#1{\,\vcenter{\openup1\jot \caja
        \ialign{\strut \hfil$\displaystyle{##}$&$
        \displaystyle{{}##}$\hfil\crcr#1\crcr}}\,}
\newif\ifdtup

\def\eqright #1\cr{\noalign{\hfill$\displaystyle{{}#1}$}}
\def\eqleft #1\cr{\noalign{\noindent$\displaystyle{{}#1}$\hfill}}

\def\oldreffmt#1{\rlap{[#1]} \hbox to 2\parindent{}}

\def\figfmt#1{\rlap{Figure {#1}} \hbox to 1in{}}

\def\sectioneq{\def\theequation{\thesection.\arabic{equation}}{\let
\holdsection=\section\def\section{\setcounter{equation}{0}\holdsection}}}%
\def\sectiontab{\def\thetable{\thesection.\arabic{table}}{\let
\holdsection=\section\def\section{\setcounter{table}{0}\holdsection}}}%
\def\sectionfig{\def\thefigure{\thesection.\arabic{figure}}{\let
\holdsection=\section\def\section{\setcounter{figure}{0}\holdsection}}}%

\newcounter{holdequation}

\def\auto{\eqno(\refstepcounter{equation}\theequation)}
\def\begineq #1\endeq{$$ \refstepcounter{equation}\eqalign{#1}\eqno
	(\theequation) $$}
\def\contlimit{\,{\hbox{$\longrightarrow$}\kern-1.8em\lower1ex
\hbox{${\scriptstyle (a\rightarrow0)}$}}\,}
\def\centeron#1#2{{\setbox0=\hbox{#1}\setbox1=\hbox{#2}\ifdim
\wd1>\wd0\kern.5\wd1\kern-.5\wd0\fi
\copy0\kern-.5\wd0\kern-.5\wd1\copy1\ifdim\wd0>\wd1
\kern.5\wd0\kern-.5\wd1\fi}}
\def\centerover#1#2{\centeron{#1}{\setbox0=\hbox{#1}\setbox
1=\hbox{#2}\raise\ht0\hbox{\raise\dp1\hbox{\copy1}}}}
\def\centerunder#1#2{\centeron{#1}{\setbox0=\hbox{#1}\setbox
1=\hbox{#2}\lower\dp0\hbox{\lower\ht1\hbox{\copy1}}}}
\def\lsim{\;\centeron{\raise.35ex\hbox{$<$}}{\lower.65ex\hbox
{$\sim$}}\;}
\def\gsim{\;\centeron{\raise.35ex\hbox{$>$}}{\lower.65ex\hbox
{$\sim$}}\;}

\def\super#1{\ifmmode \hbox{\textsuper{#1}}\else\textsuper{#1}\fi}
\def\textsuper#1{\newcount\holdspacefactor\holdspacefactor=\spacefactor
$^{#1}$\spacefactor=\holdspacefactor}

\def\supercite{\def\cite{\newcite}}

\def\newcite#1{\super{\newcount\citenumber\citenumber=0\getcite#1,@, }}
\def\getcite#1,{\advance\citenumber by1
\ifnum\citenumber=1
\ref{#1}\let\next=\getcite\else\ifx#1@\let\next=\relax
\else ,\ref{#1}\let\next=\getcite\fi\fi\next}
\def\nskip#1{\vglue-\baselineskip\vglue#1\vglue-\parskip\noindent}

\def\upon #1/#2 {{\textstyle{#1\over #2}}}
\relax

\def\m{m}

\def\Gamhat{{\widehat \Gamma}}

\def\Qbar{{\bar Q}}

\def\S{{\rm S}}
\def\P{{\rm P}}

\supercite
\def\bibitem{\item\label}

\documentstyle{article}
\font\tenbf=cmbx10
\font\tenrm=cmr10
\font\tenit=cmti10
\font\elevenbf=cmbx10 scaled\magstep 1
\font\elevenrm=cmr10 scaled\magstep 1
\font\elevenit=cmti10 scaled\magstep 1

\font\ninerm=cmr9

\renewenvironment{thebibliography}[1]
 { \elevenrm
   \begin{list}{\arabic{enumi}.}
    {\usecounter{enumi} \setlength{\parsep}{0pt}
     \setlength{\itemsep}{3pt} \settowidth{\labelwidth}{#1.}
     \sloppy
    }}{\end{list}}
\begin{document}
\nskip{-2.5truecm} \rightline{
\vbox{\elevenrm
\halign{&#\hfil\cr
&ANL-HEP-CP-92-109\cr
&November 1992\cr}
}}\nskip {1.95truecm}%

\begin{center}{{\tenbf RIGOROUS QCD PREDICTIONS\\
               \vglue 3pt
               FOR DECAYS OF P-WAVE QUARKONIA\footnotemark\\}
\vglue 1.0cm
{\tenrm GEOFFREY T. BODWIN \\}
\baselineskip=13pt
{\tenit High Energy Physics Division, Argonne National Laboratory,
Argonne, IL 60439, USA\\}
\vglue 0.3cm
{\tenrm ERIC BRAATEN\\}
{\tenit Department of Physics and Astronomy, Northwestern University,
Evanston, IL 60208, USA\\}
\vglue 0.3cm
{\tenrm G. PETER LEPAGE\\}
{\tenit Newman Laboratory of Nuclear Studies, Cornell University,
Ithaca, New York 14853, USA\\}
\vglue 0.8cm
{\tenrm ABSTRACT}}
\end{center}
\vglue 0.3cm
{\rightskip=3pc
 \leftskip=3pc
 \tenrm\baselineskip=12pt
 \noindent
\footnotetext{\ninerm
Talk presented by G.
Bodwin at DPF92, Fermilab, Batavia, Illinois, November 10--14, 1992}%
We present a new factorization theorem for the decay rates of P-wave
states of heavy quarkonia.  Infrared logarithms that had appeared in
previous perturbative calculations of P-wave decays are absorbed into
a quantity that is related to the amplitude for the heavy quark and
antiquark to be in a relative color-octet S-wave state. We predict
all of the light-hadronic and electromagnetic decays rates of the
$\chi_c$ and $h_c$ states in terms of two phenomenological parameters.
\vglue 0.6cm}
\baselineskip=14pt
\elevenrm
The annihilation of the heavy $Q\overline Q$ pair in quarkonium is a
short-distance process, occurring when the $Q$ and $\overline Q$ are
within $O(1/M_Q)$ of each other. If all of the interactions associated
with the annihilation were to occur at distance scales set by $M_Q$,
then one could, because of asymptotic freedom, calculate the decay rate
in a perturbation series in $\alpha_s(M_Q)$. Unfortunately, quark-gluon
interactions invalidate this simple scenario, since the associated
infrared (IR) and collinear singularities imply the presence of
long-range interactions.  Nevertheless, for the annihilation of S-wave
states a relatively simple picture emerges. If one neglects the
relative velocity $\vec v$ of the $Q$ and $\overline Q$,
keeping the leading term in an expansion in powers of $v/c$,
then the annihilation cross section for a meson $m$ with total spin $S$
factors into a short-distance piece times a long-distance piece:
$$
	{
\Gamma \left(\m(^{2S+1}\S) \rightarrow X \right) \; = \;
{G_1(\m)}\; {\Gamhat_1
\left( Q \Qbar (^{2S+1}\S) \rightarrow X \right)}.
}
\auto\label{factS}
$$
The quantity $\hat\Gamma_1$ is the short distance piece. It is the
(on-shell) parton-level annihilation cross section, and asymptotic
freedom allows its computation in perturbative QCD. The perturbation
series for the annihilation of an on-shell $Q\overline Q$ in an S~wave
is well behaved:  final-state IR and collinear divergences cancel
according to the KLN theorem, and initial-state IR divergences cancel
because the meson is a color singlet. $G_1$ is the long-distance piece,
which contains all of the nonperturbative effects. It is analogous to a
parton distribution. $G_1$ is proportional to the probability to find
the $Q$ and $\overline Q$ at the same point:
$$
G_1(\m)\approx (3/2\pi)|R_{\m S}(0)|^2/ M_Q^2,
\auto
$$
where $R_{\m S}(0)$ is the nonrelativistic radial wave function at the origin.

One might guess that a factorization formula similar to
Eq.~(\ref{factS}) would hold for P-wave decays as well.  However, in
this case the situation is more complicated. Owing to the angular
dependence of the wave function, the leading term in the $v/c$ expansion
vanishes upon integration over the angular orientation of $\vec v$.
Consequently, the annihilation cross section for P~waves is suppressed
by $v^2/c^2$ relative to S-wave annihilation. The first subleading term
in the $v/c$ expansion yields, in perturbation theory, a contribution
that is proportional to the {\elevenit derivative} of the wave function
at the origin.

Now, the first subleading term in the $v/c$ expansion measures the color
currents of the $Q$ and $\overline Q$, rather than their color charges.
Thus, for the subleading term, infrared divergences need not cancel, even
though the quarkonium has no net color charge. In fact, the perturbation
series for $Q\overline Q$ annihilation on shell in a P~wave contains a
logarithmic IR divergence. This is a clear signal that long distance
effects are present and that the use of perturbative QCD is not valid.
Past (nonrigorous) treatments of P-wave decays have invoked the
confinement scale or the binding energy as an IR cutoff---but with no
fundamental justification.

The structure of these IR divergences has a simple physical
interpretation.  The divergences arise when the P-wave color-singlet
state converts to an S-wave color-octet state through the emission of
a soft gluon.  Then the $Q$ and $\overline Q$ annihilate from the
S-wave state. The soft-gluon transition costs a factor $v^2/c^2$.
However, S-wave annihilation is enhanced by a factor $c^2/v^2$ relative
to P-wave annihilation, so the color-octet process is competitive with
direct color-singlet P-wave annihilation.

By taking into account the fact that the IR divergences are associated
with the color-octet mechanism, one can write a new factorization
theorem for P-wave decays:\cite{bbl}
$${
\Gamma \left(\m(^{2S+1}\P) \rightarrow X \right) \; = \;
{H_1(\m)}\; {\Gamhat_1
\left( Q \Qbar (^{2S+1}\P) \rightarrow X \right)}\;
+ \;  {H_8(\m)}\; {\Gamhat_8
\left( Q \Qbar (^{2S+1}\S) \rightarrow X \right)}.
}
\auto\label{factP}
$$
The first term in Eq.~(\ref{factP}) corresponds to the naive
(color-singlet) factorization picture. The second term is new and gives
the contribution of the color-octet mechanism. The $\hat\Gamma$'s are
the parton-level cross sections for on-shell $Q\overline Q$
annihilation, except that the IR divergent part of $\hat\Gamma_1$ is
extracted and put into $H_8$. The precise way in which this is done is
the factorization prescription. $H_1$ is proportional to the derivative
of the P-wave color-singlet $Q\bar Q$~wave function at the origin:
$$
{
H_1(\m) \; \approx \; (9/2\pi) |R^\prime_{\m P}(0)|^2 /M_Q^4.
}
\auto\label{hsinglet}
$$
$H_8$ is related to the amplitude to find the $Q$ and $\overline Q$ in a
relative color-octet S-wave state.  Since $H_8$ contains information
about the $Q\overline Q g$ Fock state, it is not simply expressible in
terms of the nonrelativistic $Q\overline Q$ wave function. In
perturbation theory
$$
{
 H_8(\m) \; \approx \; {16 \over 27 \beta_0}
\ln  \left( {\alpha_s(\epsilon_\m) \over
\alpha_s(M_Q)} \right) \, H_1(\m) \;
\sim \; {16\over 27\pi}\alpha_s\ln \left({M_Q\over \epsilon_\m}\right)\,
H_1(\m),
}
\auto\label{hoctet}
$$
where $\epsilon_\m$ is the binding energy.
The logarithm signals that Eq.~(\ref{hoctet}) is not trustworthy as an
estimate of $H_8$.  However, by comparing it with the IR divergent
parts of previous calculations of P-wave decay\cite{barbieri-et-al},
one can extract the $\hat\Gamma$'s. Both $H_1$ and $H_8$ have precise
definitions in terms of operator matrix elements, so they can, in
principle, be measured in lattice simulations, or they can simply be
treated as phenomenological parameters.

In general, $H_1$ and $H_8$ depend on the total angular momentum $J$ and
the total spin $S$ of the P-wave state.  However, if one describes the
heavy $Q\overline Q$ system in terms of a low-energy effective
Lagrangian, then the spin-dependent terms are suppressed by powers of
$1/M_Q$, that is, by powers of $v/c$.  Thus, to leading order in $v/c$,
we can take $H_1$ and $H_8$ to be independent of $S$ and $J$.
Corrections to this give terms of order $v^2/c^2$ in the decay rates,
where $v^2/c^2\approx 20\%$ for charmonium.

The factorization formula Eq.~(\ref{factP}) combined with the
leading-order expressions for the $\hat\Gamma$'s gives
$$
H_1={45\over 16\pi}{\Gamma(\chi_{c2}\rightarrow
\hbox{LH})-\Gamma(\chi_{c1}\rightarrow \hbox{LH})\over
\alpha_s^2(M_c)}, \qquad
H_8={1\over \pi}{\Gamma(\chi_{c1}\rightarrow
\hbox{LH})\over
\alpha_s^2(M_c)},
\auto
$$
where ``LH'' denotes light hadrons. Using the Particle Data
Group\cite{pdg} values for the branching ratios and the recent E760
data\cite{e760} for the $\chi_{c1}$ and $\chi_{c2}$ total widths, we
find that $H_1=15.3\pm 6.6\; \hbox{MeV}\;$ and $H_8=3.2\pm 1.4\;
\hbox{MeV}\;$. The quoted error is the experimental error, which
includes the uncertainty in $\alpha_s$, combined in quadrature with our
estimate of the theoretical uncertainty, which includes $v^2/c^2$
corrections and higher-order perturbative QCD corrections. Given $H_1$
and $H_8$ and the leading-order expressions for the $\hat\Gamma$'s, we
can use Eq.~(\ref{factP}) to compute the partial widths for the decays
of the $\chi_{c0}$ and $h_c$ into light hadrons,
the decay of $h_c$ into $\gamma$ plus light hadrons, and the decays of
$\chi_{c0}$ and $\chi_{c2}$ into two $\gamma$'s. There is also a simple
relationship between the radiative decay rates of the
$P$-states\cite{mcclary-beyers}, which is correct to leading order in
$v/c$:
$$
\Gamma( ^1\P_1 \rightarrow \gamma \; ^1\S_0 )/E_\gamma^3(11)
\; \simeq \;
\Gamma( ^3\P_J \rightarrow \gamma \; ^3\S_1 )/E_\gamma^3(3J),
\;\; J = 0, 1, 2,
\auto\label{eqptos}
$$
where $E_\gamma (11)$ and $E_\gamma (3J)$ are the energies of the
$\gamma$'s in the singlet and triplet decays, respectively.
Eq.~(\ref{eqptos}) is well satisfied for the $\chi_{c1}$ and
$\chi_{c2}$. We use it to obtain predictions for the radiative decay
widths of the $\chi_{c0}$ and $h_c$. Our predictions for the $\chi_{c0}$
are
$$
\eqalign{
\omit $\Gamma(\chi_{c0})=(5\pm 2)\;\hbox{MeV},\qquad$ \hfill
&B(\chi_{c0}\rightarrow
\hbox{LH})=(98\pm 1)\%, \cr
\omit $B(\chi_{c0}\rightarrow \gamma+J/\psi)=(2\pm 1)\%,
\qquad$ \hfill&B(\chi_{c0}\rightarrow
\gamma\gamma)=(7\pm 4)\times 10^{-4}. \cr
}
\auto
$$
The predictions for the $\chi_{c0}$ total width and branching fraction
into $\gamma +J/\psi$ differ significantly from the accepted values of
$14\pm 5$~MeV and $(0.66\pm 0.18)\%$, respectively. More precise data on
the $\chi_{c0}$ would provide useful tests of the QCD predictions. Our
prediction for the branching fraction of the $\chi_{c2}$ into two
$\gamma$'s is
$$
B(\chi_{c2}\rightarrow \gamma\gamma)=(4\pm 2)\times 10^{-4}.
\auto
$$
This is somewhat smaller than the old Particle Data Group\cite{pdg}
value of $(11\pm 6)\times 10^{-4}$.  However, a new E760
measurement\cite{e760-2}
yields a branching fraction into two $\gamma$'s of $(1.7\pm 0.6)\times
10^{-4}$. Our predictions for the $h_c$ are
$$
\eqalign{
\omit $\Gamma(h_c)=(1.0\pm 0.2)\;\hbox{MeV},\qquad$ \hfill
&B(h_c\rightarrow \hbox{LH})=(52\pm 11)\%, \cr
\omit $B(h_c\rightarrow \eta_c+\gamma)=(46\pm 11)\%, \qquad$ \hfill
&B(h_c\rightarrow \gamma+\hbox{LH})=(2\pm 1)\%. \cr
}
\auto
$$
The prediction for the $h_c$ total width is consistent with the upper
bound of 1.1~MeV obtained recently by the E760 collaboration.\cite{e760-3}
We predict a significant rate for $h_c$ into $\gamma$ plus
light hadrons.  The hard $\gamma$ recoiling against a jet
plus soft hadrons could be a distinctive signature for this decay.  A
large component of the error in all of these predictions arises from the
theoretical uncertainty.  It could be reduced considerably by making a
complete next-to-leading-order calculation of the $\hat\Gamma$'s.

One can treat quarkonium production processes using techniques that are
very similar to those that we have described for quarkonium decay
processes.  Possible applications include photoproduction,
leptoproduction, and hadron-hadron production.  The application to
production of $\chi_c$ states in $B$-meson decay can be found in
Ref.~\ref{b-decay}. The nonperturbative quantities $G_1$ and $H_1$,
which appeared in quarkonium decay, appear in the color-singlet
production processes as well. However, color-octet production involves a
new nonperturbative quantity $H'_8$.  Whereas $H_8$ is analogous to a
parton distribution function, $H'_8$ is analogous to a fragmentation
function.  The two quantities are related by crossing, but that
relationship is a simple one only in lowest-order perturbation theory.
Consequently, one must determine $H'_8$ from experiment or extract it
from a lattice calculation.

This work was supported in part by the U.S. Department of Energy,
Division of High Energy Physics, under Contract W-31-109-ENG-38 and
under Grant DE-FG02-91-ER40684, and by the National Science Foundation.
\vglue 0.5cm
{\elevenbf\noindent References \hfil}
\vglue 0.4cm

\end{document}